# Dipole Incompatibility Related Artifacts in Quantitative Susceptibility Mapping

Liangdong Zhou, Jae Kyu Choi, Youngwook Kee, Yi Wang*, and Jin Keun Seo

*Abstract*— **Artifacts in quantitative susceptibility mapping (QSM) are analyzed to establish an optimal design criterion for QSM inversion algorithms. The magnetic field data is decomposed into two parts, dipole compatible and incompatible parts. The dipole compatible part generates the desired non-streaking solution. The dipole incompatible part, including noise, discretization error, and anisotropy sources, generates artifacts defined by a wave propagator with the B0 field direction as the time axis. Granular noise causes streaking and contiguous error causes both streaking and shadow artifacts. As an example, truncation-based QSM algorithms are inherently prone to streaking artifacts because of the deviation of the dipole kernel during truncation. Through numerical phantom and imaging experiments, several QSM methods, including a Bayesian regularization method and the k-space truncation methods, were compared to show that the Bayesian approach with a structural prior provides advantages in terms of both image quality and quantitative measures.**

## I. INTRODUCTION

A tissue's molecular electron cloud becomes magnetized in an MR scanner's main B0 field to the degree characterized by tissue magnetic susceptibility $\chi$. Tissue magnetization induces its own local magnetic field, which affects the Larmor precessing frequency of the bulk water protons during MRI signal generation. Quantitative susceptibility mapping (QSM) determines the magnetic susceptibility distribution from the measured relative difference field (RDF) $b$, the local field scaled by B0, according to the dipole field model, $b = d * \chi$ [1, 2]. Given that physiological and pathological processes alter tissues' magnetic susceptibilities, QSM is useful for studying many diseases, including Parkinson's disease, Multiple Sclerosis, Alzheimer's disease, and hemorrhage [3-10].

The inverse problem of imaging $\chi$ from the measured RDF $b$ is ill-posed due to the zero cone $\Gamma_0$ of the dipole kernel in the Fourier domain [11-15]. Given a RDF data, there are many possible susceptibility solutions, and a simple kernel division causes large errors shown as streaking artifacts in the reconstructed susceptibility images [12, 16]. To deal with the ill-posedness in $\Gamma_0$, the Bayesian approach has been formulated using various prior information in susceptibility reconstruction [2, 17-21]. This includes assumptions of the susceptibility map having a piecewise constant [17, 18], smooth susceptibility or smooth susceptibility gradient [22], a sparse susceptibility gradient [11, 22, 23], or morphological consistency [22, 24-27]. These regularization methods can significantly reduce streaking artifacts, and image quality and quantification accuracy continue to be improved. [19, 28].

There are recent mathematical investigations on QSM reconstruction, with two analytical methods proposed in [29] and [30]. In [29], streaking artifacts were reduced using iterative analytic continuation near the zero cone, which produce acceptable reconstruction for noise-free field data but a poor image for noisy data. In [30], the propagation of singularity was carefully analyzed using a wavefront set, and the streaking artifacts were reduced by enforcing smoothness near the zero cone. However, the results of [30] can be biased and highly dependent on the selection of the smoothness operator.

This paper provides a mathematical understanding for artifacts in QSM. The measured field data is decomposed into a dipole-compatible part and a dipole-incompatible part, including noises, discretization error, and anisotropy sources. We show that an artifact-free solution can be obtained from the dipole-compatible part only. The dipole-incompatible part leads to artifacts that follow the wave propagation characteristics. To provide a guideline for future algorithm development, we compared four existing QSM reconstruction methods using numerical phantom data, simulated brain data, and *in vivo* brain data. The results suggest that by incorporating tissue structural priors into optimization in image space, such as MEDI [27], we can compute an optimal minimal streaking solution.

This paper is organized as follows: in Section II, we provide our main theoretical contributions (II.B, II.C, II.D) with some preliminaries on QSM inversion (II.A). Specifically, Theorem II.2 on a streaking solution with an incompatible field, Theorem II.3 on streaking with k-space truncation, and Proposition II.2 on convergence for a non-streaking solution are the core of Section II. In Section III, we demonstrate our mathematical findings in numerous experiments. In Section IV, we establish our QSM algorithm design criterion. In Section V, we discuss

The work of L. Zhou and J. K. Seo was supported by NRF grant 2015R1A5A1009350. *Asterisk indicates corresponding author.*

L. Zhou and J. K. Seo are with the Department of Computational Science and Engineering, Yonsei University, Seoul 03722, Korea (e-mail: zhould1990@hotmail.com; seoj@yonsei.ac.kr).

J. K. Choi is with the Institute of Natural Sciences, Shanghai Jiao Tong University, Shanghai 200240, China (e-mail: jaycjk@sjtu.edu.cn).

Y. Kee is with the Department of Radiology, Weill Cornell Medical College, New York, NY 10021, US (e-mail: yok2013@med.cornell.edu).

*Y. Wang is with the Department of Biomedical Engineering, Cornell University, Ithaca, NY 14853 USA, and also with the Department of Radiology, Weill Cornell Medical College, New York, NY 10065 USA, (e-mail: yiwang@med.cornell.edu).

practical issues on the implementation of our findings. Finally, we conclude in Section VI with a summary.

## II. THEORY

### A. Inverse Problem in QSM

Susceptibility distribution $\chi$ is visualized by solving the following deconvolution problem for $\chi$ [20, 25, 31]:
$$b(\mathbf{r}) = \lim_{\varepsilon \searrow 0} \int_{|\mathbf{r}-\mathbf{r}'|>\varepsilon} \frac{2(z-z')^2 - (x-x')^2 - (y-y')^2}{4\pi|\mathbf{r}-\mathbf{r}'|^5} \chi(\mathbf{r}') d\mathbf{r}' \quad (1)$$
where $b$ is the measurable field data, $\mathbf{r} = (x, y, z)$ and $\mathbf{r}' = (x', y', z')$ denote position vectors in $\mathbb{R}^3$, and the integral here is understood as the principal value [32]. This inverse problem is ill-posed, because the Fourier transform of the integral kernel $d(\mathbf{r}) = (2z^2 - x^2 - y^2)/(4\pi|\mathbf{r}|^5)$ is singular in the sense that
$$D(\mathbf{k}) = \frac{1}{3} - \frac{k_z^2}{|\mathbf{k}|^2} = 0 \text{ in } \Gamma_0 \quad (2)$$
where $D$ denotes the Fourier transform of $d$, i.e., $\mathcal{F}(d)$, $\mathbf{k} = (k_x, k_y, k_z) \in \mathbb{R}^3$ and $\Gamma_0 = \{\mathbf{k} \in \mathbb{R}^3 : k_x^2 + k_y^2 - 2k_z^2 = 0\}$. The ill-posedness caused by the zero cone $\Gamma_0$ in the frequency domain produces artifacts in the reconstructed image unless the measured field data $b$ lies in proper range space [28].

Note that $\chi$ in (1) is the solution to the following partial differential equation (PDE)
$$p(\nabla)\chi := \left(-\frac{1}{3}\Delta + \frac{\partial^2}{\partial z^2}\right)\chi = -\Delta b, \quad (3)$$
which is derived from Maxwell's equations and the MRI phase measurements of local field perturbations [2, 28, 33]. Here, $\Delta$ denotes the Laplacian operator. The fundamental solution of $p(\nabla)$ denoted as $E(\mathbf{r})$ is given by
$$E(\mathbf{r}) = \begin{cases} \frac{3}{4\pi\sqrt{z^2 - 2(x^2+y^2)}} & \text{if } \mathbf{r} \in \Upsilon \\ 0 & \text{otherwise.} \end{cases} \quad (4)$$
where $\Upsilon := \{\mathbf{r} \in \mathbb{R}^3 : 2(x^2 + y^2) < z^2\}$.

Consider the Fourier transform of the convolution problem (1); we have the following relation in Fourier domain:
$$B(\mathbf{k}) = \mathcal{F}(b)(\mathbf{k}) = D(\mathbf{k})X(\mathbf{k}) \text{ for } \mathbf{k} \in \mathbb{R}^3, \quad (5)$$
where $X$ is the Fourier transform of $\chi$. The major source of artifacts is the mismatch between the measured field data and the model (5); $\mathcal{F}(b) \neq 0$ on $\Gamma_0$, whereas $DX = 0$ on $\Gamma_0$. This mismatch is unavoidable due to noise, discretization error and anisotropy sources [28, 34].

### B. Artifact-free Solution with Dipole-Compatible Data

Since the measured field data does not match with the dipole model (5), we decompose the measured field data $b$ into
$$b = b_\diamond + b_v, \quad (6)$$
where $b_\diamond$ is the component of $b$ that can be fitted to the dipole model (5) and $b_v$ is the collection of components of $b$ that deviate from the dipole model (5) caused by noise, discretization error, and anisotropy sources.

For a systematic study, we need to appropriately choose the spaces to which $b$ and $b_v$ have to belong. Let $C_0^\infty(\mathbb{R}^3)$ denote the space of test functions, smooth and compactly supported functions on $\mathbb{R}^3$; and let $\mathcal{D}'(\mathbb{R}^3)$ denote the space of distributions in $\mathbb{R}^3$, the set of continuous linear functionals on $C_0^\infty(\mathbb{R}^3)$. The support of $u \in \mathcal{D}'(\mathbb{R}^3)$ denoted as $\text{supp}(u)$ is the smallest closed subset of $\mathbb{R}^3$ with the property that $u(\mathbf{r}) \equiv 0$ if $\mathbf{r} \notin \text{supp}(u)$. Throughout this paper, we assume that $b$ lies in $\mathcal{E}'(\mathbb{R}^3)$, the space of compactly supported distributions, and $b_\diamond$ lies in the corresponding compatible space defined as
$$\mathcal{E}'_\diamond(\mathbb{R}^3) := \left\{ u \in \mathcal{E}'(\mathbb{R}^3) : \sup_{\mathbf{k} \in \mathbb{R}^3 \setminus \Gamma_0} \left| \frac{\mathcal{F}(u)(\mathbf{k})}{P(\mathbf{k})} \right| < \infty \right\}, \quad (7)$$
where $P(\mathbf{k})$ is the characteristic polynomial corresponding to the differential operator $p(\nabla)$ in (3)
$$P(\mathbf{k}) := \frac{4\pi^2}{3}\left(k_x^2 + k_y^2 - 2k_z^2\right). \quad (8)$$
The following dipole compatible data will help us clarify the field data that is consistent with the dipole model.

**Definition II.1** (Dipole compatible field data) *Suppose that $b \in \mathcal{E}'(\mathbb{R}^3)$ is decomposed into $b_\diamond$ and $b_v$ as in (6). Let $B_\diamond$ be the Fourier transform of $b_\diamond$, i.e., $B_\diamond = \mathcal{F}(b_\diamond)$. Then, the field data $b_\diamond \in \mathcal{E}'_\diamond(\mathbb{R}^3)$ is called dipole compatible field data if it satisfies the dipole compatibility condition*
$$B_\diamond(\mathbf{k}) = D(\mathbf{k})X_\diamond(\mathbf{k}), \quad \mathbf{k} \in \mathbb{R}^3 \quad (9)$$
*for some $X_\diamond = \mathcal{F}(\chi_\diamond) \in \mathcal{E}'(\mathbb{R}^3)$. The field data $b_v$ is dipole incompatible as it cannot be fitted to the dipole model (5).*

Any compactly supported measured field data $b$ can be decomposed in the way of (6), i.e., a dipole compatible field $b_\diamond$ and a dipole incompatible field $b_v$ ($b_v$ can be zero). Let $\chi_\diamond$ and $\chi_v$ be the susceptibility maps obtained from $b_\diamond$ and $b_v$, respectively, by deconvolution. Then, we have
$$\chi = \chi_\diamond + \chi_v. \quad (10)$$
Note that $\chi_\diamond$ associated with the compatible data $b_\diamond$ is the target solution that is compactly supported on the region of interest (ROI) $\Omega$. It follows from (3), (4), and (9) that $\chi_\diamond$ satisfies the following partial differential equation:
$$\begin{cases} p(\nabla)\chi_\diamond = -\Delta b_\diamond & \text{in } \mathbb{R}^3, \\ \chi_\diamond = 0 & \text{in } \mathbb{R}^3 \setminus \Omega, \end{cases} \quad (11)$$
and can be represented as
$$\chi_\diamond(\mathbf{r}) = -\int_\Omega E(\mathbf{r} - \mathbf{r}') \Delta b_\diamond(\mathbf{r}') d\mathbf{r}'. \quad (12)$$
According to [28], the map $\chi_\diamond$ in (12) associated with the dipole compatible field data $b_\diamond$ is artifact-free.

**Theorem II.1** (Artifact-free solution [28]) *If $b_\diamond \in \mathcal{E}'_\diamond(\mathbb{R}^3)$ satisfies the dipole compatibility condition in (9), then there exists the unique artifact-free solution $\chi_\diamond \in \mathcal{E}'(\mathbb{R}^3)$ satisfying (11). The solution $\chi_\diamond$ in Fourier domain can be expressed as*
$$X_\diamond = \begin{cases} \frac{4\pi^2 |\mathbf{k}|^2 B_\diamond(\mathbf{k})}{P(\mathbf{k})} & \text{if } \mathbf{k} \notin \Gamma_0, \\ -\frac{9k_z}{4}\frac{\partial B_\diamond}{\partial k_z}(\mathbf{k}) & \text{if } \mathbf{k} \in \Gamma_0 \setminus \{\mathbf{0}\}. \end{cases} \quad (13)$$

The above theorem shows that we can obtain an artifact-free solution with the dipole compatible data even though the QSM inversion problem is ill-posed.

### C. Artifact with Dipole-Incompatible Data

The source of an artifact is, therefore, the dipole-incompatible field data that consists of noise, discretization error, and anisotropy sources. As shown in (11), the artifact-free solution $\chi_\diamond$ from the compatible data is compactly supported in the region $\Omega$; however, the map $\chi_v$ associated with the incompatible field propagates to the region outside of $\Omega$ and presents as an artifact. Hence, given the solution $\chi$ from the

measured field data $b$, the artifact $\chi_v$ from the dipole-incompatible part $b_v$ can be expressed as

$$\chi_v = \begin{cases} \chi(r) - \chi_\diamond(r) & \text{for } r \in \Omega, \\ \chi(r) & \text{for } r \in \mathbb{R}^3 \backslash \Omega. \end{cases} \quad (14)$$

The following theorem shows that the map $\chi_v$ follows the wave propagation characteristics.

**Theorem II.2** (Artifact solution) *For the incompatible field data $b_v$ in (6) with $b \in \mathcal{E}'(\mathbb{R}^3)$ being compactly supported in $\Omega$, there exists the unique $\chi_v \in \mathcal{D}'(\mathbb{R}^3)$ satisfying*

$$\begin{cases} p(\nabla)\chi_v = -\Delta b_v & \text{in } \mathbb{R}^3, \\ \chi_v = \chi & \text{in } \mathbb{R}^3 \backslash \Omega. \end{cases} \quad (15)$$

*Moreover, $\chi_v(r)$, $r \in \mathbb{R}^3$, can be represented as*

$$\chi_v(r) = -\int_\Omega E(r - r')\Delta b_v(r') dr'. \quad (16)$$

The artifact caused by the dipole-incompatible field data is defined by a wave propagator in (16) with the z variable as time. In theory, we can remove this artifact by identifying the dipole-incompatible field data $b_v$ and solving (15). It should be noted that the distribution of $b_v$ has affects the shape of the artifact. Generally, the artifact $\chi_v$ in (16) presents as a streaking artifact with granular noise $b_v$, and both streaking and shadow artifacts with contiguous error $b_v$.

The following proposition tells us that granular $b_v$ can be uniquely identified from a given $\chi$ in $\mathbb{R}^3 \backslash \Omega$ if it's in a form of multiple point sources distribution.

**Proposition II.1** *Suppose that $b \in \mathcal{E}'(\mathbb{R}^3)$ with $\text{supp}(b) \subseteq \Omega$ is decomposed as (6) where the dipole-incompatible field $b_v$ is of the form*

$$b_v(x) = \sum_{j=1}^N c_j \delta(r - r_j) \quad c_j \in \mathbb{R}, \quad (17)$$

*with distinct source points $r_1, \cdots, r_N \in \text{supp}(b)$. Then, $b_v \in \mathcal{E}'(\mathbb{R}^3)$ can be uniquely determined from the information of $\chi \in \mathcal{D}'(\mathbb{R}^3)$ in the exterior domain $\mathbb{R}^3 \backslash \Omega$ and (15).*

The proofs for Theorem II.2 and Proposition II.1 can be found in the Appendix. Theorem II.2 is the existence, uniqueness, and representation formula of the solution of PDE (15); and Proposition II.1 is the unique continuation of streaking $\chi_v$ from the exterior domain $\mathbb{R}^3 \backslash \Omega$ to $\Omega$ by using (15).

According to Proposition II.1, it is possible to identify the granular $b_v$ from $\chi$ in $\mathbb{R}^3 \backslash \Omega$ theoretically.

*D. Streaking Artifact with k-Space Truncation*

So far, we have focused on the field data – dipole compatible (Section II.B) and incompatible fields (Section II.C) – and showed that it is the dipole incompatible field data that causes artifacts. In this section, we consider the effects of information loss near $\Gamma_0$ with k-space truncation. Therefore, this section is related to QSM algorithms based on the direct inversion of (5) using k-space truncation such as TKD [12], where we show it is the loss of information or the deviation from the dipole kernel in k-space that causes another type of streaking artifact in QSM.

Many QSM methods rely on the direct inversion of (5) in the Fourier domain. With (5), $\chi$ can be obtained formally by the formula $\chi = \mathcal{F}^{-1}(B/D)$, provided that $B/D$ is properly defined near the zero cone $\Gamma_0$ as a limit [28]. Unfortunately, in the presence of dipole incompatible field data $b_v$, the fraction $B/D$ is not well defined on the zero cone $\Gamma_0$, because there exist $k \in \Gamma_0$ such that $B(k)/D(k) = \infty$. Hence, for a given measured field data, we should discard the information of $B/D$ on a neighborhood of the zero cone defined by

$$\Gamma_\epsilon \coloneqq \{k \in \mathbb{R}^3 : \text{dist}(k, \Gamma_0) < \tfrac{\epsilon}{2}\} \quad (18)$$

where $\epsilon > 0$ is related to the truncation level in Fourier domain, and $\text{dist}(k, \Gamma_0)$ denotes the distance between $k$ and $\Gamma_0$. To compensate the information loss in $\Gamma_\epsilon$, most previous methods use certain regularization with the fidelity of the following approximation [11, 22, 23, 26]:

$$\chi \approx \mathcal{F}^{-1}(B/D(1 - \eta_{\Gamma_\epsilon})) \quad (19)$$

where $\eta_{\Gamma_\epsilon}$ is the characteristic function of the set $\Gamma_\epsilon$:

$$\eta_{\Gamma_\epsilon} \coloneqq \begin{cases} 1 & \text{if } k \in \Gamma_\epsilon \\ 0 & \text{if } k \in \mathbb{R}^3 \backslash \Gamma_\epsilon. \end{cases} \quad (20)$$

The information loss of $(B/D)\eta_{\Gamma_\epsilon}$ in (19) causes artifacts in the reconstructed image. It turns out that these artifacts are different from the artifacts $\chi_v$ caused $b_v$. In theory, since $B = B_\diamond + B_v$, it must be

$$\chi_\diamond = \lim_{\epsilon \searrow 0} \left[ \mathcal{F}^{-1}\left( B/D(1 - \eta_{\Gamma_\epsilon}) \right) - \chi_v \right]. \quad (21)$$

The following theorem explains that the streaking artifacts can appear due to data loss near $\Gamma_0$ even with dipole-compatible field data, and are closely related to the singular fiber of $b$. The singular fiber of $b$ at $r$, denoted by $\sum_r(b)$, is defined as follows [35]: we say that $k \notin \sum_r(b)$ if and only if there exists $\varphi \in C_0^\infty(\mathbb{R}^3)$ with $\varphi(r) \neq 0$ and an open conic set $V$ containing $k$ such that

$$\sup_{k \in V}(1 + |k|)^N |\mathcal{F}(\varphi b)(k)| < \infty \quad \text{for all } N \in \mathbb{N}.$$

**Theorem II.3** (Streaking with k-space truncation) *Let $B_\diamond = \mathcal{F}(b_\diamond)$ satisfy the compatibility condition (9). Suppose that for $\epsilon > 0$ we have*

$$\chi_\epsilon = \mathcal{F}^{-1}\left( B_\diamond/D(1 - \eta_{\Gamma_\epsilon}) \right). \quad (22)$$

*Then, the artifacts of $\chi_\epsilon$ due to the information loss on $\Gamma_\epsilon$ can be explained by using the following wavefront set:*

$$\{(t\nabla P(k) + r, k): k \in \sum_r(b_\diamond) \cap (\Gamma_0 \backslash \{0\}), t \in \mathbb{R}\}. \quad (23)$$

It is easy to check that if $k \in \Gamma_0 \backslash \{0\}$, then $t\nabla P(k) \in \partial Y \backslash \{0\}$ for every $t \neq 0$. Hence, the artifacts due to data loss follow the wavefront set given in (23), propagating along $\partial Y$, the singular support of $E$, starting from the singular support of $b_\diamond$. This implies that the direct inversion of (19) with incompatible data has two different types of artifacts caused by the dipole-incompatible field and data loss $(B/D)\eta_{\Gamma_\epsilon}$. The following proposition shows that the data loss related artifacts in (22) can be reduced as the truncation goes to zero.

**Proposition II.2** (Convergence for artifact-free solution) *Using the same settings as Theorem II.3, there exists unique artifact-free $\chi_\diamond \in \mathcal{E}'(\mathbb{R}^3)$. The artifact-free solution $\chi_\diamond$ is the limit of $\chi_\epsilon$ as follows*

$$\lim_{\epsilon \searrow 0} \chi_\epsilon = \lim_{\epsilon \searrow 0} \mathcal{F}^{-1}\left( B_\diamond/D(1 - \eta_{\Gamma_\epsilon}) \right) = \chi_\diamond. \quad (24)$$

The proofs for Theorem II.3 and Proposition II.2 can be found in the Appendix. Please note that the convergence in (24) is equivalent to (21) and also consistent with the artifact-free solution (13) in Theorem II.1. Also note that (22) is not the standard TKD [12] because standard TKD fills the susceptibility near $\Gamma_0$ with the division of filed data by truncation level. Therefore, it is considered as a special case of TKD because (22) fills the susceptibility near $\Gamma_0$ with zero. Hence, the streaking artifacts in both (22) and TKD follow the same wavefront set characteristics as in (23).

### III. EXPERIMENTAL RESULTS

In this section, we demonstrate the mathematical statements made in Section II with numerical phantom experiments. Precisely, we verify 1) the artifact-free solution with compatible filed data in Theorem II.1; 2) the solution with streaking and shadow artifacts with incompatible data in Theorem II.2; 3) streaking with truncation in Theorem II.3 and the convergence in Proposition II.2.

A 3D numerical phantom with size $256 \times 256 \times 256$ was constructed with a piecewise constant susceptibility distribution. Four balls were placed inside the homogeneous spherical ROI. The background susceptibility was set to 0. The susceptibility values for the balls were 0.2, 0.45, 0.85 and 1, and the susceptibility in the ROI save for the balls was 0.7. The RDF data was generated by the relation (5) and the inverse Fourier transform. The true susceptibility and RDF without noise are shown in Figure 1 (a) and (b).

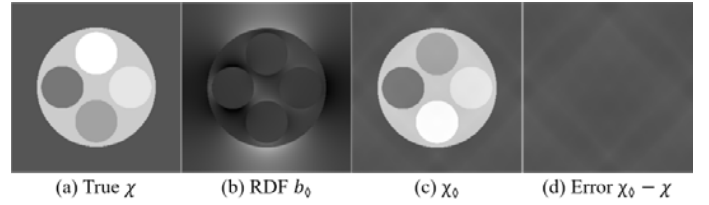

Fig.1 Sagittal image of (a) true $\chi$; (b) RDF $b_\diamond$ generated from the true $\chi$ in (a); (c) the non-streaking solution from (13), and (d) the error of the non-streaking solution. All the RDF $b$ and $b_\diamond$ for the simulated phantom are displayed in the window level $[-0.3, 1]$, and the susceptibility image in $[-0.5, 1]$.

#### A. Artifact-Free Solution with Compatible Field Data

The artifact-free solution in Theorem II.1 was reconstructed using (13) with the simulated compatible data shown in Figure 1 (b). Figure 1 (c) shows the results of the artifact-free solution and (d) shows the reconstruction error. There are no prominent artifacts in the reconstructed image, but there are minor artifacts due to computation with the discretized model.

#### B. Streaking and Shadow with Incompatible Field Data

To demonstrate Theorem II.2, two types of incompatible field, granular noise field $b_v$ and bulk error field $b_v$, were separately added into the simulated RDF data in Figure 1 (b). Since (15) cannot easily be solved in image space, we performed reconstruction using an inverse crime with the truncated k-space division (TKD) [12] (truncation level is 0.05). Figure 2 (a) shows the field data with granular noise field (red arrows), (b) is the granular noise $b_v$ in (a), (c) is the Laplacian of the granular $b_v$, (d) is the reconstructed image with field data in (a), and (e) is the error in (d). We see that (e) has only streaking artifacts (yellow dotted arrows). Figure 2 (f) displays the field data $b$ with contiguous error, (g) is the contiguous error $b_v = x^2 + y^2 + z^2$ (sphere with radius 0.35 indicated by red arrow). $b_v$ has a constant Laplacian $\Delta b_v = 6$ in the sphere shown in (h). (i) shows the reconstructed image with field data in (f), and (j) is the error of (i). Both streaking and shadow artifacts (yellow dotted arrows) are seen in (j).

Figure 2 clearly demonstrates that the incompatible field data $b_v$ causes artifacts following (16). Incompatible field data always produces streaking artifacts as shown in Figure 2 (e) and (j). The distribution of $b_v$ affects the shape of artifacts and contiguous error introduces additional shadow artifacts as shown in Figure 2 (j). The shadow appears when the waves from different directions are superposed and cannot cancel each other out. It should be mentioned that in Figure 8, we show by a decomposition approach that the waves from different directions balance out only with compatible field data.

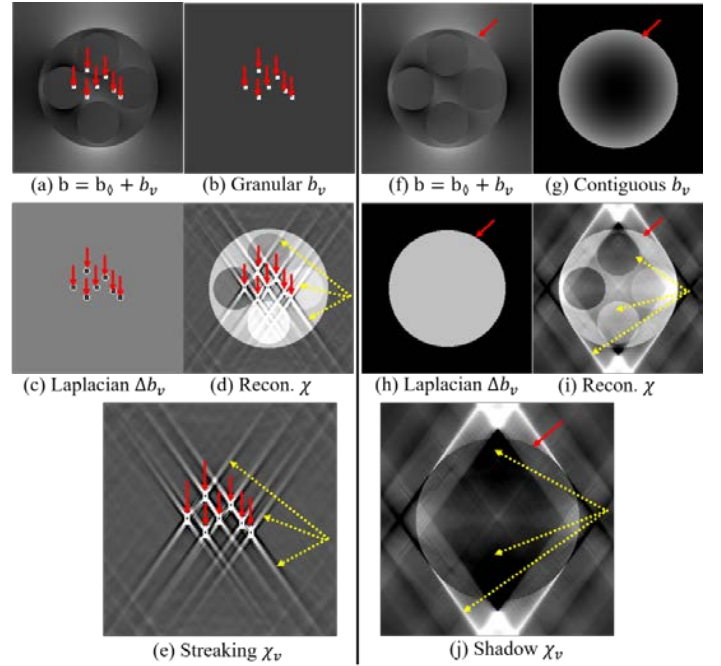

Fig.2 Demonstration of artifacts from incompatible field data. (a) the field data $b$ with granular noise (red arrows); (b) the granular noise $b_v$ in (a); (c) the Laplacian of $b_v$ in (b); (d) the reconstruction with field data in (a); (e) the streaking artifacts (yellow arrows) in (d). (f) field data $b$ with contiguous error $b_v$ (red arrow); (g) the bulk error $b_v$ in (f); (h) the Laplacian of contiguous error in (e), $\Delta b_v$, which is uniform in the sphere; (i) the reconstruction using field data in (f); and (j) the streaking and shadow artifacts (yellow arrows) in (i). To clearly see the image contrast, we set the windows for images in (c), (g), and (h) as $[-0.5, 0.5]$, $[0, 0.2]$, and $[0, 8]$, respectively.

#### C. Streaking Artifact Caused by k-Space Truncation

The data loss related streaking artifact in Theorem II.3 and the convergence in Proposition II. 2 were verified using various $\epsilon$ in (24). The $\epsilon$ was chosen such that
$$\epsilon_i = \sup\{\tau: |D| < \theta_i \text{ in } \Gamma_\tau\} \quad (25)$$
with $\theta_i = 1/2^i, i = 2, 3, \ldots, 15$.

As shown in Figure 3, with the compatible data in Figure 1 (b), the artifact due to the data loss in $\Gamma_\epsilon$, $\chi - \chi_\epsilon = \chi * \mathcal{F}^{-1}(\eta_{\Gamma_\epsilon})$ decreases as $\epsilon$ decreases where we pick up $i = 5, 8, 12$ in (25). Theoretically, as $\epsilon$ goes to zero the error $\chi - \chi_\epsilon$ vanishes. In the practical computation with discrete images, it is inevitable to lose the $B_\diamond/D$ information near $\Gamma_0$. This result implies that the artifacts due to the missing data near $\Gamma_0$ can appear in practical implementation even when $B_\phi$ satisfies the compatibility condition. Figure 3 (a) and (d) show that the streaking artifacts due to data loss propagate following the characteristics of the wave front set in (23) of Theorem II.3. The results also show that the data loss related artifacts in Figure 3 are different from the artifacts caused by the incompatible field data in Figure 2. Generally, the truncation-based methods introduce both kinds of artifacts shown in Figures 2 and 3 since the measured field data is in the form of (6).

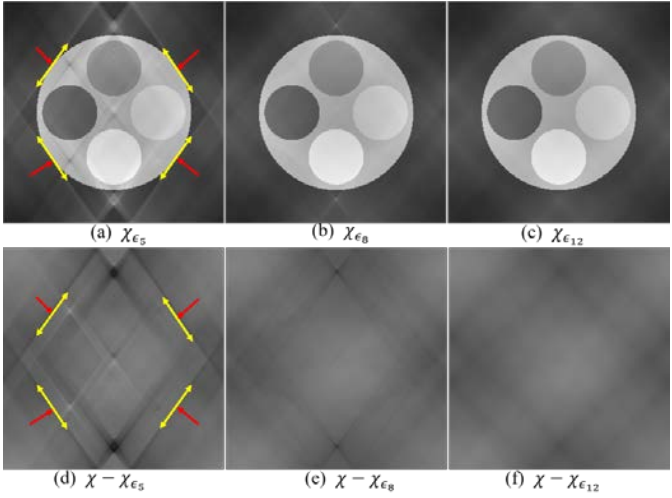

Fig.3 Sagittal image of (a)-(c) $\chi_{\epsilon_i} = \mathcal{F}^{-1}\left(B_\phi/D\left(1-\eta_{\Gamma_{\epsilon_i}}\right)\right)$ with data loss related artifacts, and (d)-(f) error $\chi - \chi_{\epsilon_i} = \chi * \mathcal{F}^{-1}(\eta_{\Gamma_{\epsilon_i}})$ which represents the artifacts due to the data loss on $\Gamma_{\epsilon_i}$. (a) and (d) show that the artifacts $\chi - \chi_{\epsilon_i}$ starts from $(r, k) \in WF(b_\phi)$ with $k \in \Gamma_0\backslash\{0\}$ (red arrows) and propagates along the straight line $r + t\nabla P(k)$, $k \in \Gamma_0$ (yellow arrows).

## IV. EXISTING METHODS COMPARISON

In this section, we establish an optimal design criterion for a QSM inversion algorithm by comparing the performances of existing methods. As discussed in Section II, the artifacts in QSM are mainly caused by the incompatibility of the field data with the dipole field, i.e., the dipole-incompatible field data and the deviation of the dipole kernel during k-space truncation.

Artifact reduction relies on probing the incompatible field data and avoiding the use of truncation in k-space. We investigated four existing QSM algorithms to assess their performance on artifact reduction. They are k-space truncation based methods: (i) the standard Truncated K-space Division (TKD) [12], (ii) the iterative Analytic Continuation method (AC) [29], (iii) fast QSM method (iLSQR) [36], and (iv) an image space regularization based method, Morphology Enabled Dipole Inversion (MEDI) [25, 27, 37].

### A. Experimental Settings

Numerical phantom data, simulated brain data, and *in vivo* brain data were used to test the four methods, TKD, AC, iLSQR, and MEDI.

*(1) Numerical phantom data (NP)*

The numerical phantom was the same as that shown in Figure 1 (a) and (b). The granular noise was added to the RDF data in the ROI to test the reduction in streaking artifacts (Figure 2 (a) and (b)). We also added contiguous error to the data (Figure 2 (f) and (g)) to test shadow artifact reduction.

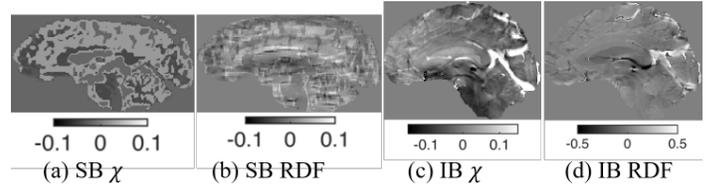

Fig. 4 Experiment settings. (a) and (b) are the true $\chi$ and RDF for simulated brain (SB); (c) and (d) are the true $\chi$ and RDF for the in vivo brain (IB).

*(2) Simulated brain data (SB)*

Complex brain MR data with matrix size $256 \times 256 \times 98$ and voxel size $[0.9375, 0.9375, 1.5]$ was generated with true susceptibility ranging from $-0.03$ to $0.19$. Gaussian noise was added to the simulated MR data. The local phase field was then calculated from the noisy MR data, therefore rendering it dipole-incompatible. This process was performed to mimic the RDF data obtaining procedure. True susceptibilities and RDF data are shown in Figure 4 (a) and (b).

*(3) In vivo brain data (IB)*

*In vivo* brain data was collected into a data matrix of size $256 \times 256 \times 146$ with voxel size $[0.9375, 0.9375, 1]$. The susceptibility distribution calculated from COSMOS [20] was set to be the ground truth for comparison of the reconstructions from the four methods. The ground truth susceptibility and RDF data are shown in Figure 4 (c) and (d).

### B. Implementation Details and Results Analysis

Several parameters were chosen for the implementation of the four methods: TKD, AC, iLSQR, and MEDI. The truncation level $\delta = 0.05$ was used in TKD for numerical phantom data and $\delta = 0.1$ was used for simulated brain and *in vivo* brain data. The implementation of AC needs the initial solution from k-space truncation as shown in (19). The truncation levels in AC were chosen as $\delta = 0.01, 0.03,$ and $0.1$ for the numerical phantom data, simulated brain data, and *in vivo* brain data, respectively. According to the original AC work [29], all results shown here were produced with 10 iterations. The iLSQR method was implemented using the conjugate gradient algorithm with error tolerance of 0.01. The MEDI algorithm was solved using the primal-dual algorithm shown recently [38] with 500 iterations. The regularization parameter $\lambda = 200$ was used for the numerical phantom data and $\lambda = 1000$ was used for simulated brain and *in vivo* brain data.

To assess the methods quantitatively, four global performance metrics for the reconstructions were evaluated: (1) root mean squared error (RMSE); (2) high frequency error norm (HFEN); (3) structure similarity index (SSIM) [39] and (4) relative total variation error (RTVE). Here RMSE is defined as $RMSE = ||\chi_{recon.} - \chi_{true}||_2 / ||\chi_{true}||_2$ and RTVE is defined as $RTVE = ||\nabla(\chi_{recon.} - \chi_{true})||_1 / ||\nabla\chi_{true}||_1$. HFEN is defined as the RMSE of the high frequency components of the images filtered with a Gaussian filter. According to the definitions of the global metrics, lower the values of RMSE, HFEN, and RTVE are better (best possible is 0), while higher values of SSIM are better (best possible is 1). HFEN and especially RMSE, show global smoothness of the reconstruction error, indicating over- or under- estimation of the reconstructions. SSIM and RTVE estimate the accuracy of structure information, such as the edges, the subdomains' interfaces of different tissues, and streaking artifacts. To assess the performance of all the methods, we define an overall relative error (OARE): OARE=RMSE + RTVE, which evaluates the overall smoothness and sharpness of the reconstruction error.

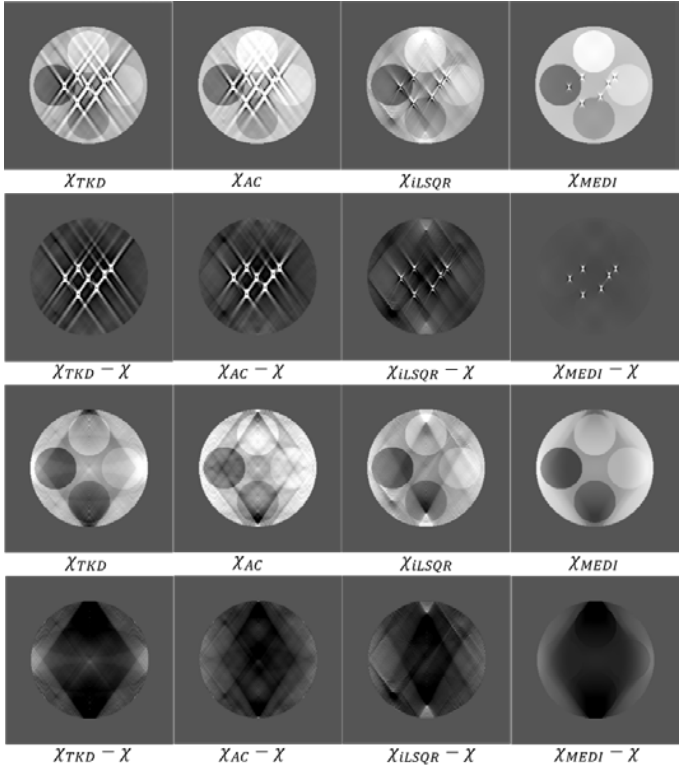

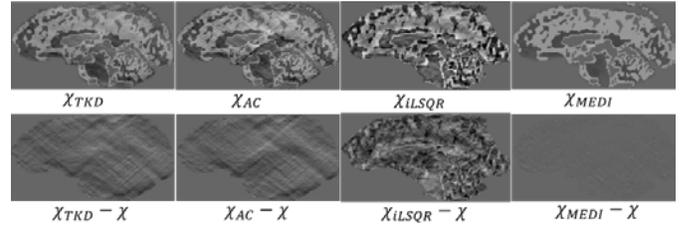

Fig. 6 Simulated brain results with simulated noisy brain data. The 1st row shows the reconstructions using TKD, AC, iLSQR and MEDI from left to right, respectively. The 2nd row shows the corresponding errors. The images are displayed in the same window $[-0.1, 0.15]$.

TABLE 2. PERFORMANCE METRICS OF FOUR METHODS WITH SIMULATED BRAIN DATA

| SB | RMSE | HFEN | SSIM | RTVE | OARE |
|---|---|---|---|---|---|
| TKD | 0.4105 | 0.2787 | 0.7196 | 0.4899 | 0.9004 |
| AC | 0.4987 | 0.3443 | 0.6567 | 2.3469 | 2.8456 |
| iLSQR | 0.9096 | 0.8952 | 0.6813 | 1.5025 | 2.4121 |
| **MEDI** | **0.1043** | **0.0683** | **0.8743** | **0.1456** | **0.2499** |

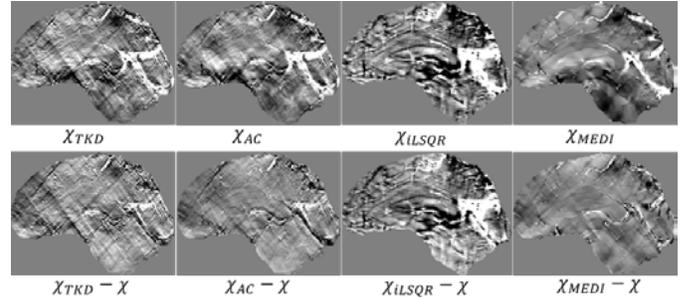

Fig. 7 *In vivo* brain results. The 1st row shows the reconstructions using TKD, AC, iLSQR and MEDI from left to right, respectively. The 2nd row shows the corresponding reconstruction errors. The susceptibilities are displayed in the window $[-0.15, 0.15]$.

TABLE 3. PERFORMANCE METRICS OF FOUR METHODS WITH *IN VIVO* BRAIN DATA

| IB | RMSE | HFEN | SSIM | RTVE | OARE |
|---|---|---|---|---|---|
| TKD | 1.0972 | 1.0784 | 0.8502 | 2.2341 | 3.3313 |
| AC | 1.0710 | 1.0889 | 0.8561 | 0.9926 | 2.0627 |
| iLSQR | 0.9912 | 1.5146 | 0.8019 | 1.1485 | 2.1397 |
| **MEDI** | **0.7896** | **0.7699** | **0.9791** | **0.5384** | **1.3283** |

Fig. 5 Numerical phantom results with granular noisy data (1st and 2nd rows) and contiguous noisy data (3rd and 4th rows). From left to right are the reconstructions and errors using TKD, AC, iLSQR, and MEDI, respectively. All the results here are displayed in the same window $[-0.5, 1]$.

TABLE 1. PERFORMANCE METRICS OF FOUR METHODS WITH NUMERICAL PHANTOM DATA

| Granular Noise | RMSE | HFEN | SSIM | RTVE | OARE |
|---|---|---|---|---|---|
| TKD | 0.3009 | 1.2922 | 0.9245 | 1.8122 | 2.1131 |
| AC | 0.1761 | 1.0402 | 0.9509 | 1.2300 | 1.4061 |
| iLSQR | 0.2513 | 0.4262 | 0.9832 | 1.138 | 1.3651 |
| **MEDI** | **0.0511** | **0.1424** | **0.9981** | **0.0756** | **0.1267** |
| Contiguous Error | RMSE | HFEN | SSIM | RTVE | OARE |
| TKD | 0.3343 | 0.3611 | 0.7761 | 1.7124 | 2.0467 |
| **AC** | **0.2689** | 0.5745 | 0.7450 | 2.7739 | 3.0428 |
| iLSQR | 0.2945 | 0.3365 | 0.8228 | 1.8890 | 2.1835 |
| MEDI | 0.2992 | **0.2101** | **0.8487** | 0.4246 | 0.7238 |

### C. Results Comparison of four QSM Algorithms

*(1) Numerical phantom (NP) results*

Figure 5 shows the results of the four methods on numerical phantom data with granular noise (1st and 2nd rows) and contiguous error (3rd and 4th rows). There are more streaking artifacts when using TKD, AC, and iLSQR than from MEDI. The performance metrics for the reconstructions are reported in Table 1, with the best values highlighted in bold type. The RMSE for all reconstructions is less than 0.4. AC has small RMSE but very large RTVE since streaking artifacts increase the value of RTVE. All four methods have an SSIM greater than 0.9 for granular noisy data, but smaller SSIMs for contiguous error. MEDI has the best HFEN, SSIM, RTVE and OARE for both cases, which means that MEDI has the smallest overall relative error, making it the best at reducing streaking artifacts among the four methods. It should be noted that based on the results in Figure 5 and Table 1, all four methods do not reduce shadow artifacts effectively. Since the shadow artifact is quite smooth in a given region, this should be reflected in the RMSE. Unfortunately, all methods over- or under- estimate the susceptibility, making it difficult to distinguish the extent of shadow artifact contribution to the RMSE.

*(2) Simulated brain (SB) results*

Figure 6 presents the results with noisy simulated brain data compared with simulated true susceptibility distribution. The error images in the second row show that MEDI produces almost the same results as the ground truth while the other three methods introduce artifacts. Performance metrics for the results are given in Table 2. MEDI performs best for all the evaluated metrics. All methods have SSIMs less than 0.8 except MEDI. AC produces severe streaking artifacts which make the RTVE large. iLSQR also shows serious reconstruction errors. The OARE implies that with simulated brain data, MEDI works best.

*(3) In vivo brain (IB) results*

Figure 7 presents the results using *in vivo* brain data, with the corresponding performance metrics presented in Table 3. Figure 7 clearly shows that all methods over- or under- estimate the susceptibility. Since there are anisotropy sources along with unknown noise in *in vivo* brain data, all reconstructions have streaking and shadow artifacts. The RMSE of all reconstructions is close to or larger than 1, with MEDI having the smallest RMSE value. The RTVE values show that TKD, AC, and iLSQR have more streaking artifacts and shape mismatches than does MEDI. The OARE indicates that MEDI outperforms the other methods in artifact reduction.

**Remarks IV.1** (QSM algorithm design criterion) *Using TKD to carry out the truncation near the zero cone causes deviation of the dipole kernel and incompatibility between field data and the dipole kernel. This incompatibility results in severe streaking artifacts. AC fills the information loss near the zero cone via analytic continuation. This may work well for dipole compatible field data. However, it cannot fill the lost data perfectly when the field data is dipole-incompatible [29]. iLSQR finds the LSQR solution of the data fitting from which the extracted streaking artifacts are deducted. From both the reconstructed images and the performance metrics, we observe that iLSQR still produces severe artifacts. As was mentioned before, the k-space truncation based methods produce artifacts caused not only by the incompatible field data but also by the data loss during truncation. This explains the poor performance of truncation based methods. MEDI is a Bayesian-based method that incorporates structural information and penalizes the sum of the susceptibility gradients across the image volume in terms of the L1-norm except for the locations where tissue edges are expected. MEDI is best at reducing streaking artifacts among these four methods; this may be due to the following two reasons: (i) since there is no truncation in MEDI, there is no loss of data, and (ii) because streaking appears along the magic angle, making it distinct from the tissue edges, tissue structure information can be used to identify a solution of minimal streaking. To reduce the shadow artifacts even further, we may need to extract the contiguous incompatible field data before performing reconstruction or carry out post processing after reconstruction. In summary, effecient reduction of the streaking artifact and the smooth structure of the shadow artifacts suggest that the development of Bayesian-based QSM reconstruction algorithms may rely on the analysis of both field data structure and artifact structure, and using proper morphological priors as regularization in image space to reduce the streaking and shadow artifacts simultaneously.*

## V. Discussion

Our mathematical analysis and experimental results demonstrate that dipole-compatible field data allows the construction of an artifact-free susceptibility distribution and that dipole-incompatible field data leads to artifacts. Granular noise causes streaking and bulk error causes both streaking and shadows.

As seen in the theories in Section II, the key idea is to examine the input field data according to its dipole compatibility. We comment on several issues regarding the implementation of dipole compatibility. First, there is a gap between the theory in continuous space and the numerical implementation in discrete space. Second, there is an ambiguity of data decomposition without given any prior information.

### A. Difficulties in Numerical Implementation of Data Decomposition in Image Domain

In Section II-C, we discussed the artifacts due to the incompatible field data and remove the source of the streaking artifacts by probing the artifact generator $b_v$ from the knowledge of streaking artifact outside of $\Omega$. In theory or with committing the inverse crime, our decomposition-based method can remove the streaking artifact. In practice, there are serious technical difficulties in its numerical implementation.

First, since the differential operator $p(\nabla)$ in (3) is a wave-type operator with the $z$ variable as the time variable, we should solve (15) with the spatial resolution $h \times h \times k$ mm$^3$ satisfying the Courant-Friedrichs-Lewy (CFL) condition $k/h \leq \sqrt{3}/2$ [40]. However, the image voxel size is pre-determined in practice, and it is difficult to match the images with numerical solutions of (15) on a voxel-by-voxel basis. Next, the numerical reconstruction of $b_v$ from the knowledge of $\chi$ in $\mathbb{R}^3 \backslash \Omega$ may not be stable, because the reconstruction method relies on the analyticity or unique continuation. Unfortunately, it's not known yet how to identify contiguous error in the field data in theory.

During the data acquisition process which obtains $b$ from the phase data of the MR signal, the data outside of the ROI is suppressed by the binary mask after the background field is removed [31, 41]. If $\Omega$ does not contain supp($b_\diamond$), then $b_\diamond$ may violate the compatibility condition because $\mathcal{F}(b_\diamond \eta_\Omega) \neq \mathcal{F}(b_\diamond)$ where $\eta_\Omega$ is the characteristic function of $\Omega$. Then, $b_v = (1 - \eta_\Omega) b_\diamond$ is not supported in $\Omega$, and our approach may not work well.

Despite its limitations, it should be noted that this is the first theoretical approach to remove streaking artifacts from the decomposition of measured field data.

### B. Ambiguity of Dipole-Incompatible Data Decomposition

Sections II-B and Section II-C revealed that the solution of (3) has a non-streaking part and streaking part corresponding to the data decomposition structure of $b = b_\diamond + b_v$. As there are difficulties in practical implementation of data decomposition, an alternative way is needed to decompose the measured data.

Let us assume $b_\diamond$ satisfies the compatibility condition (9). Decompose $b_\diamond$ into incompatible parts as the following

$$b_\diamond = b_1 + b_2 + \cdots + b_N, \qquad (26)$$

where $b_n(r)$, $n = 1, \cdots, N$ are incompatible components of $b_\diamond$. Decomposition (26) can always be done (See Example V.1). Then, there exist $\chi^n, n = 1, 2, 3, N$ such that

$$b_n(\mathbf{r}) = d * \chi^n(\mathbf{r}), \quad n = 1, 2, \cdots, N. \quad (27)$$

Since $b_\diamond$ is dipole-compatible, the corresponding reconstruction $\chi_\diamond$ will be non-streaking according to Theorem II.1. However, individual reconstruction $\chi^n$ corresponding to the incompatible data $b_n$ could have artifacts according to (16). It follows (26) and (27) that

$$\chi_\diamond = \sum_{n=1}^{N} \chi^n. \quad (28)$$

This implies that the compatible data $b_\diamond$ can be decomposed into incompatible components $b_n, n = 1, \cdots, N$ and the artifacts in $\chi^n, n = 1, \cdots, N$ will cancel out to get an artifact-free solution $\chi_\diamond$.

**Example V.1** *In the first row of Figure 8, we decompose the domain $\Omega$ into 4 subdomains such that $\Omega = \cup_{n=1}^{4} \Omega_n$ with $\Omega_n \cap \Omega_m = \emptyset$ if $n \neq m$. Under this setting, we define $b_n(\mathbf{r}) = b\eta_{\Omega_n}, n = 1, 2, 3, 4$. Then, it follows from (26) and (27) that*

$$b_\diamond(\mathbf{r}) = \sum_{n=1}^{4} b(\mathbf{r})\eta_{\Omega_n}(\mathbf{r}) = d * \sum_{n=1}^{4} \chi^n(\mathbf{r}) = d * \chi_\diamond(\mathbf{r}). \quad (29)$$

*From Fourier transform and the convolution theorem, (29) can be rewritten as*

$$\sum_{n=1}^{4} B_\diamond * \mathcal{F}(\eta_{\Omega_n}) = D \sum_{n=1}^{4} X^n. \quad (30)$$

*Note that the convolution operator is nonlocal. Therefore, the compatibility condition is violated for each $b_n$:*

$$B_n = B_\diamond * \mathcal{F}(\eta_{\Omega_n}) \neq 0 = DX^n \quad \text{on } \Gamma_0. \quad (31)$$

*Hence, the reconstruction of $\chi^n$ with $b_n$ will have artifacts due to data incompatibility (the third and fourth rows of Figure 8). The k-space data $B_n$ is nonzero on $\Gamma_0$, therefore, it is dipole-incompatible. However, the summation of all decomposed k-space data is zero on $\Gamma_0$ (the second row of Figure 8). This is why the reconstructions using decomposed data contain serious streaking and shadow artifacts, but the summation of all the reconstructions from decomposed data has no artifacts. This example also explains that the contiguous error $b_v$ in (16) causes both streaking and shadow artifacts. If the field data is compatible, then the superposition of wave propagation from different directions cancel out to make an artifact-free solution.*

**Remark V.1** (Ambiguity of incompatible data decomposition) *The Example V.1 explains that the compatible data can be decomposed into incompatible parts. In practice, the measured data is incompatible. Following the notation in Example IV.1, if the measured data $b$ is*

$$b = b_1 + b_2 + b_3 = b_\diamond - b_4, \quad (32)$$

*it may not be possible to find the compatible part $b_\diamond$ and incompatible part $b_4$ from $b$, since $b_4$ and $b$ occupy entirely different regions and contain different information (See Figure 8). Indeed, there are also infinitely many ways to decompose compatible field data $b_\diamond$into compatible parts. The combination of compatible data and incompatible data leads to incompatible data. Hence, unique decomposition of measured incompatible field data into (6) is impossible without using any prior information.*

This ambiguity of incompatible data decomposition again suggests the use of prior information or regularization for unique data decomposition. This also matches with the QSM algorithm design criterion established in Remark IV.1, suggesting that effective data decomposition relies on enforcing prior information in a proper way[24-27].

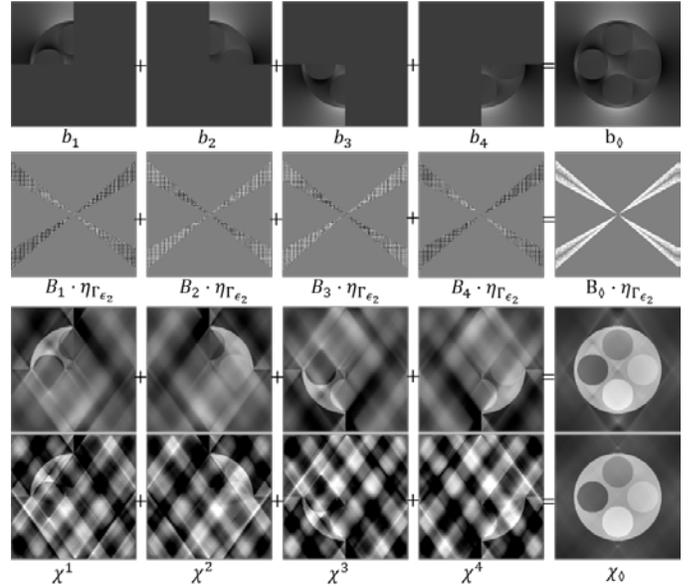

Fig.8 Sagittal image of b and its compositions $b_n, n = 1,2,3,4$ and the corresponding reconstructions $\chi^n, n = 1,2,3,4$ with $\mathcal{F}(b)$ satisfy the compatibility condition. The first row shows the data decomposition and the second row shows the corresponding k-space data constrained on $\Gamma_{\epsilon_2}$ in the window level $[-5, 5]$. The third and fourth rows are reconstructions with two different $\epsilon$. Apparently, the decomposed k-space data in the second row do not satisfy the compatibility condition since they are nonzero on $\Gamma_0$. This incompatibility introduces the artifacts (both streaking and shading) in the reconstructions.

## VI. CONCLUSION

The mathematical features in QSM were investigated by the wave operator acting on a susceptibility map and the Laplacian operator on the magnetic field. The dipole-compatible field data generates the non-streaking solution, even though the inverse problem is ill-posed. The dipole-incompatible field data caused by noise, discretization error and anisotropy sources results in artifacts, particularly streaking artifacts from granular noise and shadow artifacts from contiguous error. Although we are not aware of an optimal method that decomposes measured field data into a dipole-compatible part and a dipole-incompatible part, we can obtain a minimal streaking solution by using MEDI, a Bayesian approach which incorporates tissue structural priors. Despite the advantages in streaking artifact removal by applying MEDI, it is still a challenge to remove the smooth shadow artifact.

Future QSM studies should address the various technical challenges to deal with the noisy $b$ on the zero cone $\Gamma_0$ as a way to improve accuracy. This requires innovation based on rigorous mathematical analysis to make the problem $B = DX$ well defined by suitably extracting the dipole-incompatible components.

## APPENDIX

### A. Proof of Theorem II.2

**Proof.** Noting that the fundamental solution $E$ of $p(\nabla)$ in (4) is locally integrable and the support of $b_v$ is included in $\Omega$, the existence of $\chi_v$ in $\mathcal{D}'(\mathbb{R}^3)$ is obvious from the representation formula in (16). The exterior condition of $\chi_v = \chi$ in $\mathbb{R}^3 \backslash \Omega$ is from (14).

For the uniqueness, it is sufficient to show that $u \equiv 0$ is the unique solution of
$$\begin{cases} p(\nabla)u = 0 & \text{in } \mathbb{R}^3, \\ u = 0 & \text{in } \mathbb{R}^3 \backslash \Omega. \end{cases}$$
Since $u$ is supported in $\Omega$ and it satisfies $p(\nabla)u = 0$ in $\mathbb{R}^3$, its Fourier transform $\mathcal{F}(u)$ is extended to an entire analytic function in the complex space $\mathbb{C}^3$ with $P(\boldsymbol{k})\mathcal{F}(u)(\boldsymbol{k}) = 0$ for all $\boldsymbol{k} = (k_x, k_y, k_z) \in \mathbb{C}^3$. Since $\mathcal{F}(u)$ is an entire analytic function which vanishes in the open set $\{\boldsymbol{k} \in \mathbb{C}^3 : P(\boldsymbol{k}) \neq 0\}$, it must be $\mathcal{F}(u) = 0$, and therefore $u \equiv 0$ by Fourier inversion theorem. This completes the proof. ∎

*B. Proof of Proposition II.1*

**Proof.** Let $\chi \in \mathbb{R}^3 \backslash \Omega$ be given. To derive a contradiction, we assume that there exist two different $b_v^1, b_v^2 \in \mathcal{E}'(\mathbb{R}^3)$ of the form (17) satisfying (15). Let $\chi_v^1$ and $\chi_v^2$ in the space of $\mathcal{D}'(\mathbb{R}^3)$ be the solutions of (15) corresponding to $b_v^1$ and $b_v^2$, respectively. Then, $u = \chi_v^1 - \chi_v^2$ satisfies
$$p(\nabla)u = -\Delta(b_v^1 - b_v^2) \quad \text{in } \mathbb{R}^3 \tag{33}$$
$$u = 0 \quad \text{in } \mathbb{R} \backslash \Omega \tag{34}$$
where $b_v^1 - b_v^2$ can be expressed in the form:
$$b_v^1(\boldsymbol{r}) - b_v^2(\boldsymbol{r}) = \sum_{j=1}^N c_j \delta(\boldsymbol{r} - \boldsymbol{r}_j). \tag{35}$$
Since $u$ is compactly supported from (34), $\mathcal{F}(u)$ is an entire analytic function in the complex space $\mathbb{C}^3$ by the Paley-Wiener-Schwartz Theorem [35]. $\mathcal{F}(u)$ is determined from the identity (33) and (35):
$$\mathcal{F}(u)(\boldsymbol{k}) = \frac{4\pi^2|\boldsymbol{k}|}{P(\boldsymbol{k})} \left( \sum_{j=1}^N c_j e^{-2\pi i \boldsymbol{k} \cdot \boldsymbol{r}_j} \right) \quad \text{for all } \boldsymbol{k} \in \mathbb{C}^3. \tag{36}$$
Because the function $\mathcal{F}(u)$ is entirely analytic, we must have $c_j = 0$ for all $j = 1, \cdots, N$, and therefore, we have
$$b_v^1 - b_v^2 = 0 \quad \text{in } \mathbb{R}^3.$$
This completes the proof. ∎

*C. Proof of Theorem II.3*

**Proof.** The reconstructed $\chi_\epsilon$ in (22) can be decomposed into the true part $\chi_\diamond$ and artifact $\chi_a = (\chi_\epsilon - \chi_\diamond)$
$$\chi_\epsilon = \chi_\diamond + (\chi_\epsilon - \chi_\diamond) = \chi_\diamond + \chi_a. \tag{37}$$
Using the convolution theorem for inverse Fourier transform, $\chi_a = (\chi_\epsilon - \chi_\diamond)$ can be represented as
$$\chi_a = -\mathcal{F}^{-1}\left(\frac{B_\diamond}{D}\eta_{\Gamma_\epsilon}\right) = -\mathcal{F}^{-1}\left(\frac{B_\diamond}{D}\right) * \mathcal{F}^{-1}(\eta_{\Gamma_\epsilon}) \tag{38}$$
Moreover, with a suitable use of L'Hôpital's rule [28], the term $B_\diamond/D \, \eta_{\Gamma_\epsilon}$ in (22) is approximated as
$$\frac{B_\diamond(\boldsymbol{k})}{D(\boldsymbol{k})} \eta_{\Gamma_\epsilon}(\boldsymbol{k}) = \left(\frac{\partial}{\partial k_z}B_\diamond(\boldsymbol{k}) / \frac{\partial}{\partial k_z}D(\boldsymbol{k}) + O(\epsilon)\right)\eta_{\Gamma_\epsilon}(\boldsymbol{k})$$
$$= \left(-\frac{9k_z}{4}\frac{\partial B_\diamond}{\partial k_z}(\boldsymbol{k}) + O(\epsilon)\right)\eta_{\Gamma_\epsilon}(\boldsymbol{k}). \tag{39}$$
It follows from (38) and (39) that the artifacts of $\chi_\epsilon$ can be expressed as
$$\chi_a = -\frac{9}{4}\mathcal{F}^{-1}\left(k_z \frac{\partial B_\diamond}{\partial k_z}\right) * \mathcal{F}^{-1}(\eta_{\Gamma_\epsilon}) + O(\epsilon)\mathcal{F}^{-1}(\eta_{\Gamma_\epsilon}). \tag{40}$$
From (40), when $\epsilon > 0$ the artifacts $\chi_a = (\chi_\epsilon - \chi)$ can be expressed as
$$\chi_a = -\frac{9}{4}\int_{\mathbb{R}^3} \mathcal{F}^{-1}(\eta_{\Gamma_\epsilon})(\boldsymbol{r} - \boldsymbol{r}')\left[b_\diamond(\boldsymbol{r}') + z'\frac{\partial b_\diamond}{\partial z'}(\boldsymbol{r}')\right]d\boldsymbol{r}'$$
$$+ O(\epsilon)\mathcal{F}^{-1}(\eta_{\Gamma_\epsilon}). \tag{41}$$
Note that $\frac{\eta_{\Gamma_\epsilon}}{\epsilon}$ behaves like Dirac delta function $\delta_{\Gamma_0}$ given by
$$\delta_{\Gamma_0}(\boldsymbol{k}) = \delta(\text{dist}(\boldsymbol{k}, \Gamma_0)).$$
Therefore, $\chi_a/\epsilon = (\chi_\epsilon - \chi_\diamond)/\epsilon$ can be approximated by
$$\chi_\#(\boldsymbol{r}) := -\frac{9}{4}\int_{\mathbb{R}^3}\mathcal{F}^{-1}(\delta_{\Gamma_0})(\boldsymbol{r} - \boldsymbol{r}')\left[b_\diamond(\boldsymbol{r}') + z'\frac{\partial b_\diamond}{\partial z'}(\boldsymbol{r}')\right]d\boldsymbol{r}', \tag{42}$$
where $\mathcal{F}^{-1}(\delta_{\Gamma_0})$ is defined in the sense of tempered distribution. Hence, the artifacts of $\chi_\epsilon$ can be explained by the wave front set of $\chi_\# \approx \chi_a/\epsilon = (\chi_\epsilon - \chi_\diamond)/\epsilon$.

The proof of the characterization of streaking artifacts of $\chi_\epsilon$ follows the work in [28]. For the sake of clarity, we briefly include the proof. The wavefront set of $\chi_\#$ is defined as
$$\text{WF}(\chi_\#) = \{(\boldsymbol{r}, \boldsymbol{k}) \in \mathbb{R}^3 \times (\mathbb{R}^3\backslash\{0\}) : \boldsymbol{k} \in \Sigma_{\boldsymbol{r}}(\chi_\#)\}.$$
Note that the wavefront set of $u * v$ satisfies
$$\text{WF}(u*v) \subseteq \{(\boldsymbol{r}+\boldsymbol{r}',\boldsymbol{k}) : (\boldsymbol{r},\boldsymbol{k}) \in \text{WF}(u) \& (\boldsymbol{r}',\boldsymbol{k}) \in \text{WF}(v)\} \tag{43}$$
for $u \in \mathcal{D}'(\mathbb{R}^3)$ and $v \in \mathcal{E}'(\mathbb{R}^3)$ [35]. The proof of (23) will be completed if we examine $\text{WF}(\mathcal{F}^{-1}(\delta_{\Gamma_0}))$. Then, because $\delta_{\Gamma_0}$ can be regarded as a density distribution on $\Gamma_0$, its wavefront set satisfies that the first variable is point $\boldsymbol{k}$ on $\Gamma_0$ and the second variable is normal vector field $t\nabla P(\boldsymbol{k}), t \neq 0$ of $\Gamma_0$:
$$\text{WF}(\delta_{\Gamma_0}) = \{(\boldsymbol{k}, t\nabla P(\boldsymbol{k})) : \boldsymbol{k} \in \Gamma_0\backslash\{0\} \& t \neq 0\}$$
$$\cup \{(0, t\nabla P(\boldsymbol{k})) : t \neq 0 \& P(\boldsymbol{k}) \geq 0\}.$$
Note that $\delta_{\Gamma_0}$ is homogeneous which enables us to determine $\text{WF}(\mathcal{F}^{-1}(\delta_{\Gamma_0}))$ from the support and the singular support of $\delta_{\Gamma_0}$ [35]. In other words, we have
$$\text{WF}(\mathcal{F}^{-1}(\delta_{\Gamma_0})) = \{(t\nabla P(\boldsymbol{k}), \boldsymbol{k}) : \boldsymbol{k} \in \Gamma_0\backslash\{0\} \& t \in \mathbb{R}\},$$
and by (43), we finally have
$$\text{WF}(\chi_\#) \subseteq \{(t\nabla P(\boldsymbol{k})+\boldsymbol{r},\boldsymbol{k}) : \boldsymbol{k} \in \Sigma_{\boldsymbol{r}}(b_\diamond) \cap (\Gamma_0\backslash\{0\}), \, t \in \mathbb{R}\}.$$
This completes the proof of Theorem II.3. ∎

*D. Proof of Proposition II.2*

**Proof.** According to Theorem II.1, $\chi_\diamond$ is non-streaking. Since we have $\lim_{\epsilon \searrow 0} \mathcal{F}^{-1}(\eta_{\Gamma_\epsilon}) = 0$ in the sense of tempered distribution. Therefore, it follows from (37) and (38) that
$$\lim_{\epsilon \searrow 0}\chi_\epsilon = \chi_\diamond + \lim_{\epsilon \searrow 0}(\chi_a) = \chi_\diamond + \chi_\diamond * \lim_{\epsilon \searrow 0}\mathcal{F}^{-1}(\eta_{\Gamma_\epsilon}) = \chi_\diamond.$$
This proves the convergence in (24). ∎